\documentclass[runningheads,a4paper]{llncs}
\usepackage{amsmath}
\usepackage{graphicx}

\usepackage{url}
\urldef{\mailsa}\path|{floriana.gargiulo, sylvie.huet}@cemagref.fr|

\title{How opinion dynamics generates group hierarchies}
\author{F. Gargiulo, S. Huet}

\institute{Cemagref, Laboratoire d'Ingenierie pour les Systèmes Complexes,\\
BP 50085, 63172 Aubiere Cedex, France\\
\mailsa\\
\url{http://www.cemagref.fr/lisc}}

\begin{document}

\maketitle

\begin{abstract}
We recently proposed a model coupling the evolution of the opinions of the individual with the local network topology. The opinion dynamics is based on the Bounded Confidence model. The social networks is based on a group concept where each individual is totally connected to the members of its group and is linked to the individuals of the other groups with a given probability. During a time step, the individual has to decide between discussing with a member of its own network and applying the opinion dynamics, or moving groups because it has an opinion far from the average opinion of its own group.  One of the main results we obtained is that the group sizes, starting from an homogenous situation can be strongly heterogeneous at the equilibrium state. This kind of heterogeneity can be identified in many real networks. In this paper we present the complete set of behaviours that this complex model can exhibit, at group level. In particular we will focus on the mechanisms that lead to the stability of the groups and the appearance of heterogeneity in sizes. 

\end{abstract}

\section{Introduction}
Each of us knows that a system can be stable at a given scale while it is not at another one. That is what we are interested in and this paper is going to describe how population, group and individual levels can differ in stability and forms. But, let's come back to the beginning presenting some assumptions of the simple simulated model of the society we work with.

Other's opinion is a source of cognitive inconsistency! That is what Festinger \cite{festinger1957Dissonance} argued adding that it is experienced as dissonance. According to him, the dissonance is a psychological discomfort or an aversive drive state that people are motivated to reduce, just as they are motivated to reduce hunger. In his balance theory, \cite{heider1946} used a similar concept and called it imbalance. More recently, \cite{MatzWood2005} showed that, as the dissonance and balance theories suggest, the disagreement from others in a group produces cognitive inconsistency and the negative states of dissonance or imbalance.

The groups are a privileged place of interaction between people and the exchange with others can lead to dissonance. They are thus at the same time the entity creating dissonance and the one reducing it. Indeed, three strategies can be chosen to reduce its dissonance created by the heterogeneity of the opinion inside its group (see \cite{MatzWood2005}): changing one's own opinion to agree with others in the group, influencing others to change their opinions, or joining a different, attitudinally more congenial group. 

The two first relates to the individual interactions which are often based on similarity and have been extensively studied in the attraction paradigm \cite{Byrne1971} and other theories on interpersonal interactions as the social judgment theory \cite{SherifHovland1961}. The third can be linked to the personal external network of the individual. Indeed, its external neighbours give it some information about the characteristics of their group and can introduce it. This stresses out the importance of the individual's social networks.

The structure of social networks has been the object of study of many different disciplines. In particular, a more quantitative investigation of the topological structure of social networks has been possible with the great availability of data of web 2.0. 
The analysis performed with this tools revealed  that social networks usually present  community based structures: analyzing networks at different scales it is possible to identify groups of persons much more interconnected among them than with the rest of society \cite{girvan2002community}. Many different algorithm have been created to identify communities on large networks \cite{2009community} and many models have been proposed to explain the mechanism leading to the formation of such underlying structure \cite{palla2007quantifying}.

Recently many works have been done regarding evolving network topology and their adaptation to the social background \cite{gross2008adaptive}: as people can influence each other to induce a change of mind, the difference of opinion on some very important topics can also lead to the breaking of a social contact. In other terms, since people prefer to be surrounded by persons sharing similar opinion (homophily) it is quite likely that  the change of opinions due to the opinion dynamics processes can lead to the change of the  network structure. An interesting analysis of the co-evolution of opinions and networks is presented in \cite{kozma2008consensus}.

Starting from the strategies to overcome inconsistency explained above, we have presented in \cite{Gargiulo1002.1896} a simulation model reproducing their main aspects in order to better understand the link between the individual choices and the organization of the society into groups. We will model the group concept as the one of community. It is based on a social network where an individual has most of its links to its own group and a minor part to the other groups. The interaction process between individuals will be modeled by one of the classical opinion dynamic model.

\section{The model}
The model presented in \cite{Gargiulo1002.1896} couples an opinion dynamics process with an adaptive network structure co-evolving with the opinion. The considered network has a typical underlying structure based on groups. We defined group as a set of individuals fully interconnected among them. From the point of view of the individuals, therefore, each agent has a link to all the other members of the same group. Moreover individuals can also be connected, with a fixed probability $p_{ext}$, to members of different communities. The resulting network therefore is composed by different complete graphs (the groups) connected among them through the links that agents establish outside the group. At the initialization time each agent select the membership in a group. When all the agents have decided their membership, the internal links are established. Secondly, each agent connects with probability $p_{ext}$ to all the agents that are not inside its own group. The initialization is completed randomly assigning to each agent a continuous opinion $\vartheta$ ranging in the interval $[0,1]$.

Agents choose between two possible dynamics at a time: they can communicate with an other agent in their neighbourhood (exchanging opinion with it and updating their opinions with the opinion of the other agent); or they can reconsider their membership in the group. The choice of being part of a group, in fact, is usually based on homophily preferences: if the average socio-cultural attitudes of a group are strongly different from the one of a member, this member could feel a certain uneasiness and decide to change groups. In this case we will consider homophily preferences based on the opinion. In a simulation step all the agents are selected and, at their turn,  they can decide with probability $p_{change}$ to change their membership in a group and with probability (1-$p_{change}$) to perform opinion dynamics. 

When an agent $i$ chooses to perform opinion dynamics, it first selects a partner, $j$, for the contact process: the partner can be selected both among the members of its own group or the agents connected through external connections. The opinion updating mechanism is the Bounded Confidence (BC) model introduced by Deffuant et al. in \cite{deffuant2000mixing} and deeply analyzed in all its aspects in \cite{ben2003bifurcations}, \cite{lorenz2007continuous}, \cite{fortunato2004universality}:

\begin{equation}
\mbox{if }|\vartheta_i(t)-\vartheta_j(t)|<\varepsilon \qquad \left\{\begin{matrix}
 \vartheta_i(t+1)= \vartheta_i(t)+\mu(\vartheta_j(t)-\vartheta_i(t)) \\
   \vartheta_j(t+1)= \vartheta_j(t)+\mu(\vartheta_i(t)-\vartheta_j(t))
\end{matrix}\right.
\end{equation}

The parameter $\varepsilon$ represents the tolerance of an agent and it is the fundamental critical parameter of the system. The parameter $\mu$, on the other side, represents the relative shift of the agents after a discussion.

On the other side, the mechanism to change groups follows the following steps: the agent looks at the distance between its opinion and the average opinion of its own group, $O_I$. If $|\vartheta_i(t)-O_I(t)|>\varepsilon$, it changes groups looking at the groups with which it already has a contact through an external link. It will connect to a chosen group with a probability:

\begin{equation}
 P_{i\rightarrow J}=\frac{1-|\vartheta_i(t)-O_J(t)|}{\sum_{J\supset j\in \mathcal{V}(i)}\left(1-|\vartheta_i(t)-O_J(t)|\right)} 
\end{equation}

where $\mathcal{V}(i)$  is the neighbourhood of node $i$.

Once an agent change groups, its connections are completely rewired: all the connections he had before (inside the previous group and outside) are canceled. It is connected to all the agents of the new group and new external connections are established with the probability $p_{ext}$ with the agents outside the new group. All the relationship are symmetrical: this means that if an agent A is linked to an agent B, the agent B is also linked to the agent A.

In previous papers (\cite{Gargiulo:arXiv0912.2821}, \cite{Gargiulo1002.1896}) we  provided a detailed description of the opinion dynamics process on this kind of group-based adaptive networks, specifying the main differences with respect to static networks. We discovered some surprising behaviours compared to the known behaviour of the Deffuant bounded confidence model(BC).
First of all, the coupled model exhibits a total consensus for an $\varepsilon$ value lower than the BC model. That is linked to the capacity of the coupled model to suppress the minor clusters positioned in the BC model on the extrema of the opinion space. In social psychology, groups are known as a source of cohesion and avoidance of the isolation. Thus, that is a very interesting fact that the introduction of groups in the BC model suppresses the isolated individuals. Thus, we observed an heterogeneity of the final size distribution of groups which is more realistic than the more homogeneous size distribution obtained with the Bounded Confidence model.

An other interesting phenomenon concernes the distribution of the group sizes: for certain values of $\varepsilon$, some groups become larger as the system evolves while other decrease in size, sometimes until containing only one individual.

Finally, from the point of view of the groups, the consensus remains for a large set of $\varepsilon$ values while, looking at the population level, there are a lot of opinion clusters. Then, each group does not simply correspond to a subpopulation exhibiting the same behaviour than the whole population. 

In this paper we focus on the obtained equilibrium and the form of the size distribution. The results presented in the following sections concern simulations with \textit{N} = 5000 individuals and \textit{G}=500 groups. Some parameters of the model are fixed ($\mu=0.5$, $p_{ext}=0.001$ and $p_{change}=0.5$) in order to study the characteristic behaviours in different ranges of the tolerance parameter $\varepsilon$. Regarding $\varepsilon$, we study the value for which the total consensus is not reached. Practically, it means we are interested in the various $\varepsilon$ lower than 0.267.

\section{Different tolerances, different equilibria}

\begin{figure}
\centering
\includegraphics[width=10cm,]{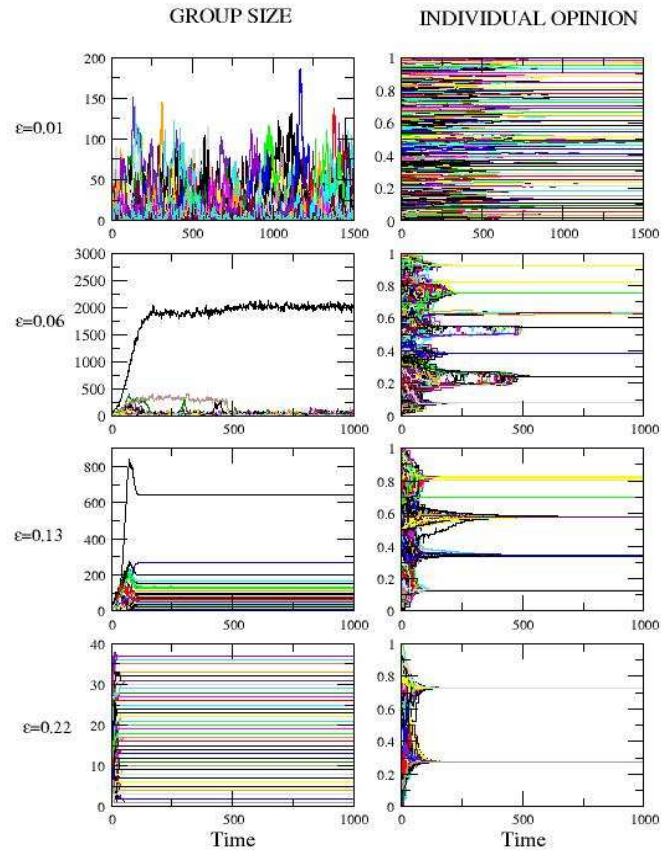}
\caption{Left column: Evolution in time of group sizes. Each line represents one group. Right column: Evolution in time of individual opinions. Each line represent an individual. Each raw corresponds to a different value of $\varepsilon$: 0.01, 0.06,0.13,0.22}
\label{1real}
\end{figure}

Fig. \ref{1real} displays the typical evolutions of group sizes and individual opinions in time for one realization of various values of $\varepsilon$. As we can observe from the right column, the opinion dynamics leads to the formation of stable opinion clusters as it happens for static networks (even if the cluster structure is not exactly the same as for the static case \cite{Gargiulo1002.1896}).

On the other side, looking to the left column of the plot, the situation regarding groups is very different: for some values of $\varepsilon$ (0.01 and 0.06) group sizes continues to vary in time. Individuals, even if their opinion is stable, continue to jump from one group to an other and this process never reaches a steady state. This is not the case for $\varepsilon=0.13$ and $\varepsilon=0.22$. for which, after a transitory phase, groups, as individuals, become stable over time also.

We can argue therefore the presence of different kinds of equilibrium, for some values of the tolerance parameter $\varepsilon$, we find stable opinions and unstable groups, above a threshold value, also the groups stabilize. To study this transition to static equilibrium we consider the relative standard deviation of the size in the \textit{T}=100 last time steps at the population level opinion equilibrium, for each of the groups:

\begin{equation}
 \frac{\sqrt{\langle S_I^2 \rangle_T - {\langle S_I \rangle_T}^2}}{\langle S_I\rangle_T}
\end{equation} 

where the symbol $\langle...\rangle_T$ indicates the average on time. 

\begin{figure}
\centering
  \includegraphics[width=8cm,clip=true]{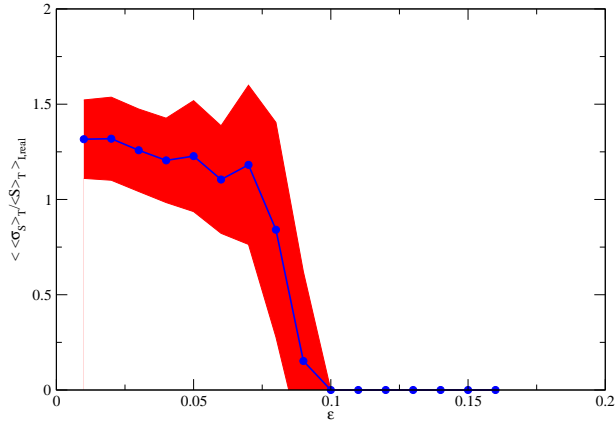}
  \caption{Relative standard deviation on time, averaged on all the groups and on 100 realizations}\label{varSize}
\end{figure}

This indicator, averaged for all the groups and on 100 independent  realizations is displayed, with its relative error, in Fig. \ref{varSize}. As we can observe from this figure, the transition to group stability can be set at $\varepsilon\sim0.1$. 

\begin{figure}
\centering
  \includegraphics[width=8cm]{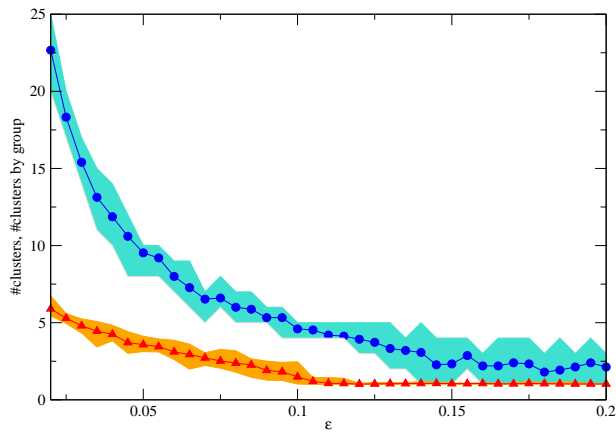}
  \caption{Number of opinion clusters by group and number of opinion cluster at the population level for various values of $\varepsilon$. The result is averaged on 30 realizations. The dotted line represents the minimum and the maximum on the realizations.}\label{nbOpClust}
\end{figure}

This result is coherent with what we observed in \cite{Gargiulo1002.1896} where we studied the percentage of individuals susceptible to change groups as a function of $\varepsilon$.  For $\varepsilon<0.1$ it is impossible to obtain the situation \emph{one opinion cluster in one group}; therefore the dynamics of groups cannot converge to a steady state.  This situation is described in Fig. \ref{nbOpClust}. First of all, we can notice that the number of opinion clusters inside a group is always lower than the number of clusters at population level. In particular, for $\varepsilon=0.1$, we observe that the average number of opinion clusters at population level is five, but each group contains only one cluster. For $\varepsilon<0.1$, therefore, the system is unstable: when there are several opinion clusters by group, some individuals are necessary far from the average opinion of the group since the distance between two opinion clusters is equal to at least $\varepsilon$. 

\begin{figure}
\centering
  \includegraphics[width=10cm,clip=true]{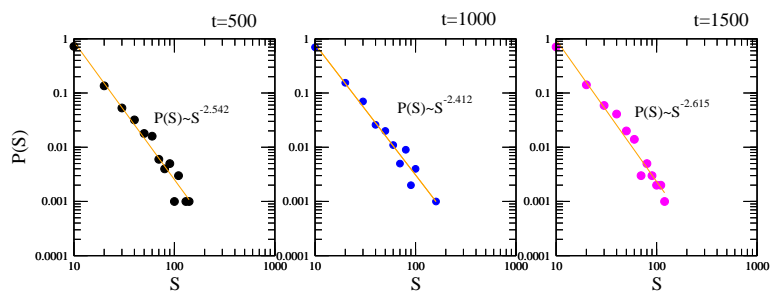}
  \caption{Group size  distribution for $\varepsilon=0.01$ at different moment (T=500, 1000,1500) of the evolution. The result is averaged on 100 realizations.}\label{sizeDistrTime}
\end{figure}

Since in the following we are going to explore some statistical properties of the size distributions, it is important to analyze in details the non-equilibrium phase. As we observed before, for $\varepsilon<0.1$ the groups are not stable. Notwithstanding this dynamical instability, if we observe the aggregate properties of the system, namely the size distribution at different times of the evolution, we can argue that the statistical properties of the model remain constant. In the Fig. \ref{sizeDistrTime} we analyzed the case $\varepsilon=0.01$ and we displayed the group size distribution at different moments of the evolution. The figure shows that the size distributions maintain over time the same shape. This confirms that the size distribution remains a good indicator of the heterogeneity of the group size in the non-equilibrium phase.  

\section{The various levels of heterogeneity}

\begin{figure}
\centering
  \includegraphics[width=10cm,clip=true]{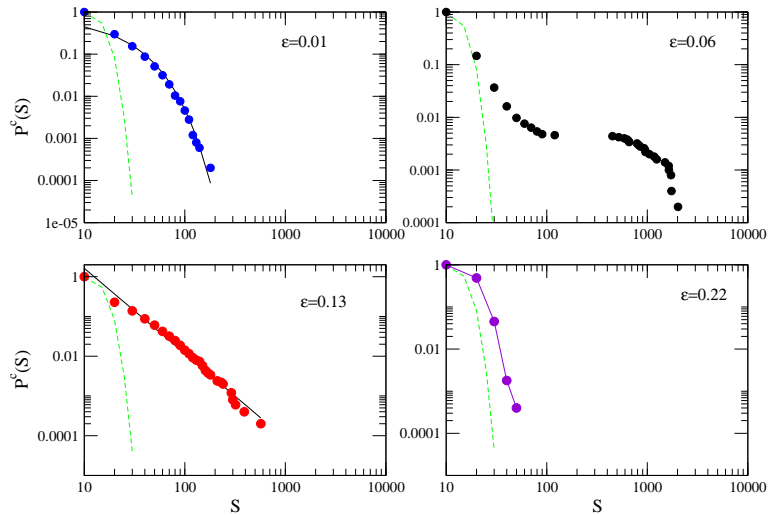}
  \caption{Cumulative size distribution at the end of the evolution for $\varepsilon=0.01,0.06,0.13,0.22$. The dotted lines represent the size distribution at the initialization. The result is averaged on 100 realizations. }\label{CumSizeDistr}
\end{figure}

From the Fig. \ref{1real} we can also extract interesting considerations regarding the distribution of group sizes. Since we use a random procedure to fill the group structures, at the initial state,  all the groups are distributed with a small deviation ($\sigma_S=3.15$) around the average value: $\langle S\rangle= N/G=10$. The average size of the larger group, at the beginning is $S_{max}=20.8\pm 1.62$. Of course the average size of the groups will remain the same during all the evolution of the system since the population $N$ and the number of groups $G$ are constant. By the way, looking at the Fig. \ref{1real} and taking into account the considerations about Fig. \ref{sizeDistrTime}, we observe that for $\varepsilon=0.01$, some bigger groups ($S\sim 100$) already appear. At $\varepsilon=0.06$ a macroscopic group containing more than 20\% is present, while all the other groups seem to be much smaller. The situation is again different at $\varepsilon=0.1$ where many intermediate sizes seem to be allowed. Finally, for $\varepsilon=0.22$ the situation seems to be similar to the initialization moment. A statistical description of these different situations is displayed in Fig. \ref{CumSizeDistr} where the cumulative size distributions for the four cases analyzed in Fig. \ref{1real} are reported. 

Fig. \ref{CumSizeDistr} provides a statistical description of the assumption we did by the observation of the typical run in Fig. \ref{1real}: for $\varepsilon=0.01$ group sizes present a distribution that decrease fast with the size and that can be fitted with an Exponential law. For $\varepsilon=0.06$ we observe a bi--modal distribution, where the large tail corresponds to the  few macroscopic groups. For $\varepsilon=0.13$ group sizes present a sort of hierarchy structure where all the sizes seem to be possible. The data, in this case,  can be fitted with a power law distribution, denoting the absence of a typical scale in the system.  Finally the size distribution for $\varepsilon=0.22$ is not strongly dissimilar to the initial one.

\begin{figure}
\centering
  \includegraphics[width=10cm,clip=ture]{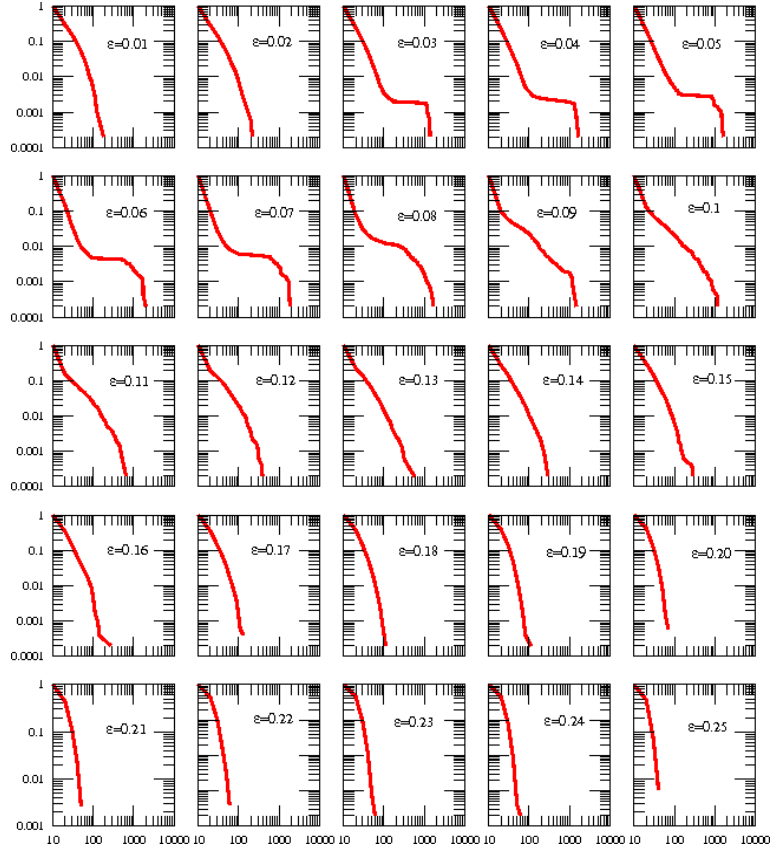}
  \caption{Cumulative size distribution at the end of the evolution for all the analyzed values of $\varepsilon$. The result is averaged on 100 realizations.}\label{sizeTutti}
\end{figure}

A more complete set of cumulative degree distributions for different values of $\varepsilon$ is given in figure \ref{sizeTutti}.

\begin{figure}
\centering
  \includegraphics[width=8cm]{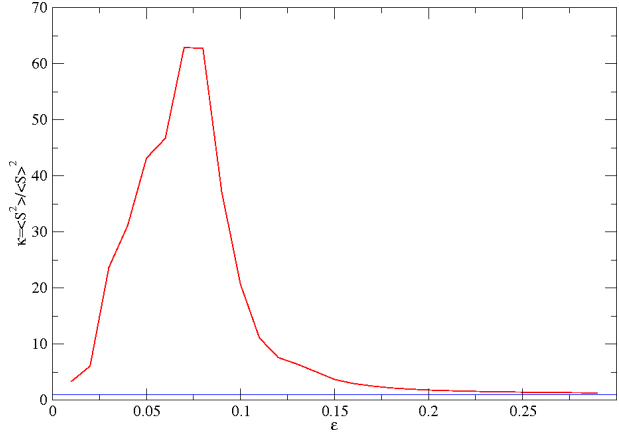}
  \caption{Heterogeneity between group sizes as a function of $\varepsilon$. The result is averaged on 100 realizations.}\label{heterogeneity}
\end{figure}

The variability of the sizes, at the final state is described from the second moment of the size distribution $\langle S^2\rangle_{I,real}$. We can provide a characterization of the heterogeneity level of the group structures introducing the parameter: $\kappa=\langle S^2\rangle/(\langle S\rangle)^2$: if $\kappa\gg1$ the variance is dominated by the second moment and therefore the sizes are strongly heterogeneous, while for $\kappa\sim1$ we have an homogeneous distribution. The Fig. \ref{heterogeneity} displayed this parameter as function of $\varepsilon$.

As we can see, the parameter $\kappa$ quickly increases for $\varepsilon>0.02$, reaches a pick for $\varepsilon\sim0.07$ and rapidly decreases. Connecting this information with  the analysis of the size distributions (Fig. \ref{sizeTutti}) we can argue that:
 for $\varepsilon\leq0.02$ the system presents an exponential size-distribution and, therefore,  it is characterized by a low value of heterogeneity.
The range $0.02<\varepsilon\leq0.1$ corresponds to the bi-modal distribution, with the two characteristics scales (small groups and few macroscopic groups). This is the phase of higher heterogeneity. The range  $0.1<\varepsilon\leq0.14$, instead, is characterized by a scale free group size distribution. Also this phase presents a high level of heterogeneity, even if lower than in the previous case. Finally the distribution returns gradually to the initial one (passing through situations very similar to the ones obtained for $\varepsilon\leq0.02$). For $\varepsilon>0.2$ the groups result to be homogeneous in size among them.

\section{An heuristic explanation of stability and heterogeneity}

The mechanism that generates heterogeneity is a hidden preferential attachment that estabilishes the rapid growth of some particular groups. The model in fact implies that an individual, when he chooses a group, considers in his selection only the groups on which he can retrieve information, namely the groups in which he has a friend (a link). Since the choice of the external links does not depend on the membership (an agent has the same probability $p_{ext}$ to select all the agents that are non member of its own group), the probability that an external link ends in a big group is higher. The fact that more links bring to the bigger groups, makes these more eligible as possible destinations when agents decide to change groups. The creation of heterogeneity is therefore an effect of this preferential attachment. The different shapes of the size distribution (exponential, bi--modal, power law) are a consequence of the fact that for different values of $\varepsilon$ the \emph{rich get richer} mechanism happens in different ways. At the  beginning, due to the random initialization, the average opinion of each group is $O_I\sim 0.5$. Therefore, on average, all the individuals with an opinion $\vartheta$ situated in the range $0.5-\varepsilon<\vartheta<0.5+\varepsilon$ are stable: since their opinion is very similar to the average opinion of the group, they are satisfied and they won't change groups. The group dynamics therefore starts from the agents with opinions further than $\varepsilon$ from the center of the opinion interval. When $\varepsilon\rightarrow 0$ the number of agent that are initially susceptible to move is higher. 

Another important point is that the average initial number of individuals in a group is: $n_I=N/G$. The average distance between two individuals at the initial point therefore is: $d=1/(n_I-1)$, since the opinion space measures $1$. When $d\geq\varepsilon$ opinion dynamics, on average, does not happen. With our choice of parameters it is always the case for $\varepsilon<0.1$.

\begin{figure}
\centering
  \includegraphics[width=12cm]{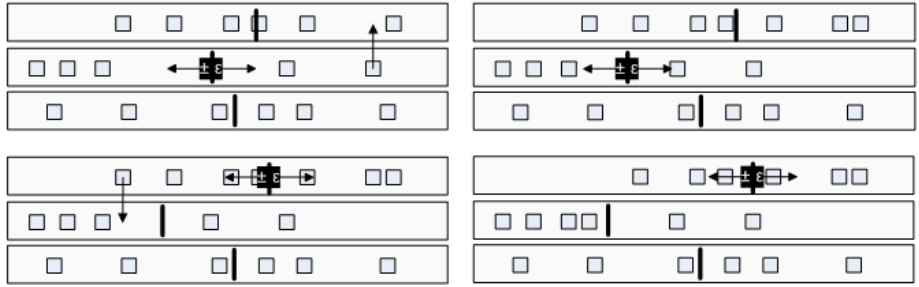}
  \caption{Group dynamics polarisation effect. Evolution of a population of 18 individuals in 3 groups. The time goes from the left to the right starting from the top. Time 0 to 3.}\label{traj1}
\end{figure}

Two mechanisms linked to the both dynamics are responsible for the heterogeneity emergence. At a first stage, the initial small unbalance due to stochasticity of the average opinion of groups, favors a polarization of the groups. The Fig \ref{traj1} shows how it occurs due to the dynamics of group change. It represents 3 groups with 6 agents (grey squares) in each. Agents with an opinion far from the average (the dark vertical line) start to move to groups with a closer $O_I$. These moves implies a polarisation of the average opinion of the group which, at its turn, leads in each group to a desertification of one opinion extreme in favour of the other.

\begin{figure}
\centering
  \includegraphics[width=12cm]{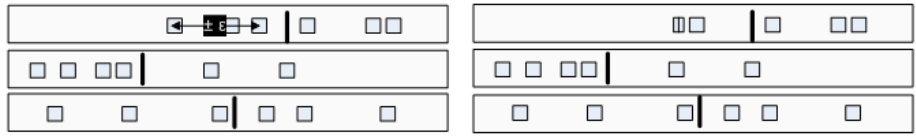}
  \caption{Opinion dynamics decreasing standard deviation effect. Evolution of a population of 18 individuals in 3 groups. The time goes from the left to the right. Time 4 and 5}\label{traj2}
\end{figure}

As people condense on one extreme, opinion dynamics happens reducing the standard deviation of the group opinion. The Fig. \ref{traj2} shows how two close (regarding $\varepsilon$) individuals influence each other leading to more cohesiveness of the group members because the minimum opinion of the group become closer to the maximum one. This standard deviation decreasing phenomenon gives to the new individual arriving in the group more chance to be satisfied and to remain in the group. 

One can see that at a given time, the number of individuals in each group can differ a lot. That is sufficient for the preferential attachment described previously to allow a larger group to become larger and larger. Depending on the number of opinion clusters, the more or less remaining central people are going to attract the more extreme opinion clusters and, if they are enough, make the group having an average opinion close to the center. That is the most frequent case. The exception is when $\varepsilon$ leads to three clusters at the population level.

This phenomenon leading to the emergence of heterogeneity does not occur completely for $\varepsilon\rightarrow 0$ because all the agents (also the central ones) move since the beginning, forbidding the polarization of group opinions. For higher values of $\varepsilon$ the polarization starts to occur, but the probability of finding a group able to gather the central opinions ( i.e. a group containing a higher number of central individual stable since the beginning) is lower, since the stability interval is very small. Just one or few groups have this characteristic and, therefore, the capacity of starting the \emph{rich get richer} mechanism. This leads to the bi--modal distribution.

Finally, as $\varepsilon$ increases, the number of groups suitable for the preferential attachment also increases, allowing the simultaneous growth of more than one macroscopic group. This is the origin of the power law distribution.

Stability is connected to the relative speed of the opinion dynamics process and of the group dynamics. As $\varepsilon\rightarrow0$ opinion dynamics became slower, never allowing the formation of \emph{one cluster in one group}. Opinion dynamics, in fact, for low values of $\varepsilon$ can happen only at population level on large time scale, without influencing the group formation. As we observed before, for higher values of $\varepsilon$ opinion dynamics start to be possible also at group level. Since the number of connections inside groups is higher, this mechanism accelerates opinion dynamics process. The combination of opinion dynamics and group dynamics bring in such cases to specification of opinions inside groups, \emph{one cluster in one group}. In intermediate cases, like for $\varepsilon=0.06$, many small groups composed of two symmetrical side clusters remains. This situation allows a continuous exchange between the central cluster and the groups in such situation.  

\section{Conclusions}
In this paper we analyzed the properties of the groups obtained by a combined dynamics of opinions and membership choices. In particular we analyzed two aspects: the different kind of equilibrium and the size distributions. 

For what concerns the equilibrium, two main situations can be identified: for $\epsilon<0.1$, even if all the opinion clusters are formed, the dynamics of groups never reaches a stable state. In this case, however, even if the individuals continue to jump from one group to the other, at the aggregate level the degree distribution results stable. For $\varepsilon\geq0.1$, instead, the system reaches an equilibrium where neither the opinions nor the groups change over time. 

Regarding the group sizes we identified four characteristic regions: low heterogeneity phase ($\varepsilon<0.02$); few groups dominance ($0.02<\varepsilon<0.1$) where macroscopic structures are formed; group hierarchy ($0.1<\varepsilon<0.2$) and group homogeneity ($\varepsilon>0.2$).

The phenomenological analysis we provided, by the way, highlights some important points: first of all, the dynamics strongly depends on the number of initial groups and, in particular from the initial number of individuals in the groups. This point is a subtle one and  a more analytical study would be needed, through a mean field analysis. 

Moreover, we tried to eliminate the preferential attachment mechanism, giving to the agents the possibility to do a rational choice between all the groups independently from the network structure. With this option, we didn't observe any heterogeneity for $\varepsilon<0.1$ but some heterogeneity remains in the other case. Also in this case, some simple mathematical models, would be needed to provide a better quantitative explanation. 

Finally, what we obtained, except for very low values of $\varepsilon$, is that the heterogeneity level of sizes decreases when $\varepsilon$ increases. The significance of such a result is not easy to catch. Indeed one can notice that the $\varepsilon$ used for group assessment and the one used for individual assessment are quite different in their meaning. They could be considered as two parameters in another model. On the one hand, smaller is $\varepsilon$, larger is the need for cohesion of the individual to recognize a group as its own group. On the other hand, larger is $\varepsilon$, more tolerant is the individual regarding the opinion difference to others. Concretely, it means that an individual more defines itself by its own opinion when $\varepsilon$ is low while it more defines itself by its group membership when $\varepsilon$ is large. One can see a kind of individualistic society versus a collective one. That is very counterintuitive is that the individualistic society leads to the emergence of large majority groups while the collective society leads to a much more homogeneous size of groups! A deeper investigation in the social psychological litterature would be necessary to assess about the realistic feature of this result.

\bibliographystyle{plain}	
\bibliography{GO_refs}
\end{document}